\documentclass[]{aipproc}

\layoutstyle{8x11double}

\begin{document}

\title[GRB Hosts are Starbursts]{Unveiling the Progenitors of Gamma-Ray
Bursts through Observations of their Host Galaxies}

\author{Ranga-Ram Chary}{
address={UCO/Lick Observatory, University of California, Santa Cruz, CA 95064},
email={rchary@ucolick.org}
}

\begin{abstract}
Analysis of the multi-wavelength broadband photometry between rest-frame ultraviolet
and near-infrared wavelengths
indicates that the extinction corrected
star-formation rates per unit stellar mass of a small sample of gamma-ray burst (GRB)
host galaxies
are higher than those of prototypical, nearby starbursts. This result, the 
confirmed detection of 
the host of GRB980703 at radio wavelengths, and the tentative evidence in favor
of a supernova light curve underlying the visible light transient associated with some GRBs
provides evidence for a connection between stellar phenomenon and GRBs.
Fitting population
synthesis models to the multiband photometry of the host galaxies
reveals the presence of a young stellar population
with age less than 50 Myr in 4 out of 6 galaxies, albeit with large uncertainties. 
Determining the age of the stellar population in a large number of GRB host galaxies
using this technique could be one of the more
reliable ways of distinguishing between the collapsar model for GRBs
and models that involve the merger of degenerate objects 
in a binary system. 
\end{abstract}

\maketitle

\begin{section}{The Starburst-GRB Connection}
Detailed analysis of the cosmic infrared background and long wavelength
galaxy counts by various groups have illustrated the large ($>$70\%) 
contribution
from infrared luminous galaxies (L$_{\rm IR}$=L(8$-$1000~$\mu$m)>10$^{11}$~L$_{\odot}$)
to the star-formation rate density at $z<3$ \cite{cha01, fra01}. 
Infrared luminous galaxies emit as much as 90\% of their bolometric luminosity
at far-infrared wavelengths indicating that dust reprocessing is extremely 
significant at z$\sim$1$-$3.
Most of these galaxies are inconspicuous at optical/UV wavelengths
except that many of them show evidence of merger activity. 
This seems to suggest that tidally induced starbursts dominate the co-moving 
star-formation rate (SFR). However, the classification of a starburst galaxy
based on its UV-determined SFR is ambiguous.
For example M82, which is not an infrared luminous galaxy, is classified as a 
starburst
galaxy based on stellar population synthesis model fits to it's optical/near-infrared 
multiband photometry.
These fits yield a young age for the stellar population, of order 10 Myr. While
it's star-formation rate is about 7~M$_{\odot}$~yr$^{-1}$, no higher than that of M51 
(the Whirlpool Galaxy),
it's specific star-formation rate i.e. the star-formation rate per unit stellar 
mass is about
an order of magnitude higher than M51, promoting it to the `starburst' galaxy 
category. 
Thus, if GRBs were indeed associated with massive stars,
almost all the hosts should be starburst galaxies with a large fraction of
these being infrared luminous objects. To test this hypothesis it is necessary to 
obtain either a measure of the
far-infrared luminosity or the extinction corrected specific SFR in the GRB host galaxies.
\begin{figure}
\resizebox{0.48\textwidth}{!}{\includegraphics{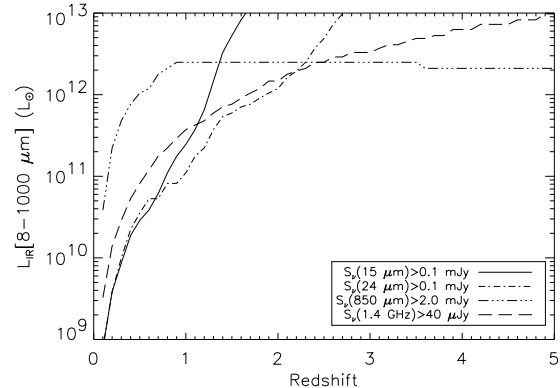}}
\caption{
Minimum infrared luminosity that can be detected by
various long wavelength observations as a function of redshift assuming the template
spectral energy distributions of \cite{cha01}.
The sensitivity values adopted correspond to the
limits of ISOCAM (15$\mu$m), SCUBA (850$\mu$m) and VLA/WSRT (21 cm).
The 24$\mu$m
sensitivity is the MIPS/SIRTF limit for a $\sim$30 minute integration.
The mid-infrared regime is the most sensitive to estimating dust-enshrouded
star-formation at $z<2$. If we ignore contributions to the IR light from an AGN,
L$_{\rm IR}=10^{10}$~L$_{\odot}$ corresponds to a star formation rate of 1.7~M$_{\odot}$/yr.
}
\label{fig:uno}
\end{figure}
Figure 1 shows the sensitivity of the deepest surveys at a variety of mid- and far-infrared
wavelengths which allow a direct determination of L$_{\rm IR}$ (see also \cite{elb02}). 
Clearly, the mid-infrared
regime (7$-$25~$\mu$m) is most sensitive to detecting infrared luminous galaxies at $z<2.5$. 
However, even the deepest surveys at these wavelengths observe only galaxies
with L$_{\rm IR}>10^{11}$~L$_{\odot}$.

This is further illustrated by the fact that of the
16 GRB host galaxies that have been observed so far
at 850~$\mu$m using the SCUBA instrument, only one of them (GRB010222)
has been detected at late-times after the afterglow has faded completely \cite{berg02}. 
The derived SFR for this object is comparable to the SFR inferred from the
radio observations of GRB980703 \cite{berg02} suggesting that
it is an ultraluminous infrared galaxy (ULIG).

The low detection rate of GRB host galaxies in the submillimeter
questions the validity of the association between star-formation
and GRBs.
However, at the typical SCUBA
sensitivity of 3$\sigma\sim$3~mJy,
only the brightest ULIGs that are more 
luminous than L$_{\rm IR}\sim$4$\times$10$^{12}$~L$_{\odot}$ can be
detected at $z>1$. These hyper luminous galaxies
comprise only a small subset of the entire ULIG population.
Since the total contribution from ULIGs to the global star-formation
history is at most 30\% \cite{cha01} the contribution
from these hyper-luminous sources is less than 10\%.

Thus, as inferred independently by \citet{ram01},
only about 10\% of GRB hosts
are likely to be detected by SCUBA at sensitivity limits of 3~mJy which is consistent with the
small detection rate mentioned above and the small number of statistics.
MIPS/SIRTF observations will be vital in deriving the far-infrared luminosity of a large number
of GRB hosts.

Fortunately, at the faint end of the infrared luminosity function, a complementary technique
which utilizes the UV-slope seems to measure the dust-obscured star-formation
reasonably well \cite{meu99}. At L$_{\rm IR}<4\times10^{11}$~L$_{\odot}$, much of the dust obscuration
is optically thin i.e. the far-infrared emission is dust reprocessed
UV emission from the same part of the galaxy.
However, in the more luminous sources the dust is concentrated so strongly that
the UV opacity is very high in the neighborhood of the star-forming regions and so {\it all}
the UV photons are re-radiated at longer wavelengths.
In this case, the UV and far-infrared observations trace the low and high opacity regions
of the same galaxy respectively and the UV-slope is only a lower limit to the amount of obscuration.

To search for signatures of a starburst in the GRB hosts using this technique, the
multiband photometry on GRB host galaxies between rest-frame ultraviolet
and near-infrared wavelengths were fit using the stellar population synthesis models
of Bruzual \& Charlot, taking into account internal dust extinction \cite{cha02}. 
The total (obscured+UV)
star-formation rate in these galaxies is not found to be unusually high (Figure 2).
Although only 2 of the 6 host galaxies with accurate photometry
have derived
L$_{{\rm IR}}>10^{11}$~L$_{\odot}$ and high resultant SFRs, the
ratio between the dust-obscured SFR and the unobscured
SFR for
the GRB hosts has an average
value of $\sim$4. This is in agreement with the redshift
dependent value of 3$-$7 that \citet{cha01} find for the ratio between the global comoving
dust-obscured
star-formation rate and the unobscured star-formation rate derived from the UV.

The near-infrared wavelengths constrain the stellar mass and age of the galaxy very well (Figure 3).
Our analysis of the stellar population age and baryonic
mass of 6 of the host galaxies with accurate photometry and redshifts reveal
that all of them have high specific star-formation rates 
compared to local starbursts (Figure 2).
The spectral energy distribution of 4 of them (GRB970228, GRB970508, GRB980613, GRB980703)
are dominated by a young (age$<$50~Myr) stellar population which
favors the collapsar model. However, 2 of them (GRB990123, GRB991208) have stellar populations
of age $\sim$200~Myr suggesting that the merger of double degenerate systems might power some of the GRBs.
\begin{figure}
\resizebox{0.48\textwidth}{!}{\includegraphics{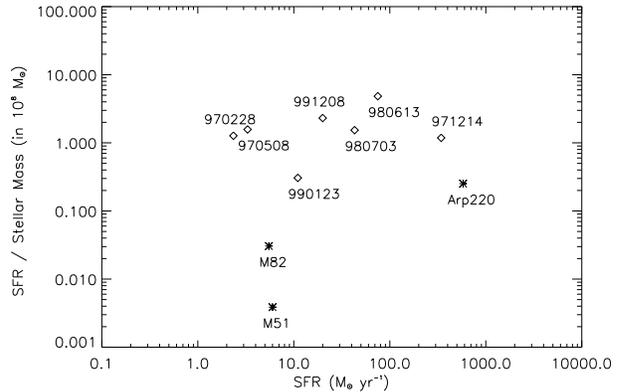}}
\caption{
Total (obscured+unobscured) star-formation rates for selected
gamma-ray burst host galaxies derived using the $\beta$-slope technique
plotted against the ratio between the star formation rates and the midpoint of the range
of stellar masses derived from fits to the multiband photometry.
The uncertainty on the GRB971214 point is quite large because of the
lack of good photometry at multiple wavelengths.
The star-formation rates for the GRB hosts are lower
limits, so the data points are likely to move higher and to the right.
Estimates of the stellar mass typically have 95\% confidence intervals that span an order of magnitude.
Also plotted are the corresponding values
for two prototypical starbursts Arp220 and M82 and the relatively quiescent Sbc galaxy M51. 
GRB hosts have star-formation rates per unit stellar mass much higher than local starbursts.
}
\label{fig:due}
\end{figure}
\begin{figure}
\resizebox{0.48\textwidth}{!}{\includegraphics{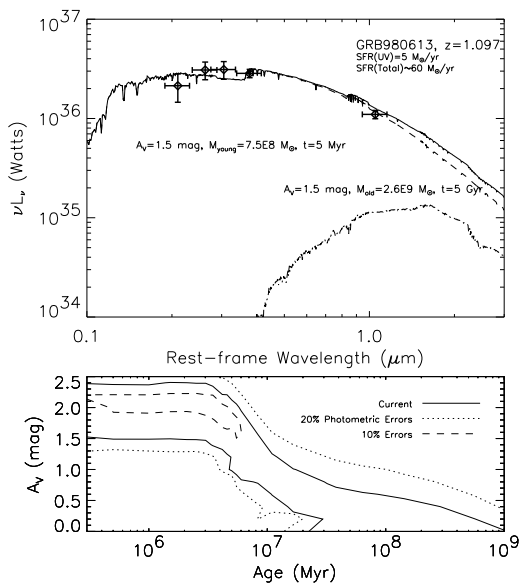}}
\caption{
Optical/near-infrared photometry for the host of GRB980703 along with a template SED
of a galaxy with an exponentially decaying starburst. The template consists of
a young stellar population fit to the optical/UV photometry and the maximal contribution from an
older stellar
population which might have formed in an initial burst at $z\sim$10 and has evolved over the 
5~Gyr between
$z=10$ and $z=1.097$. The solid black line is the sum of the two components.
The luminosity of many GRB host galaxies are
dominated by a young (10-50~Myr) stellar component which favors
an origin of GRBs from collapsars. GRB host galaxies whose SEDs are 
dominated by an older stellar component
would favor alternate mechanisms such as merging neutron stars. The lower
plot shows the range of values for the extinction and age which result in SEDs
that are within 2$\sigma$ of the data points at B,V,R,I and K. The different contours illustrate
the variation in the parameters with photometric uncertainty.
}
\label{fig:f3}
\end{figure}
\end{section}

\begin{section}{Future Work}
Various techniques have been suggested to constrain the progenitors of GRBs: studying the
environments of GRBs through spectral analysis of
the X-ray afterglow \cite{gal01}, searching for signatures
in the light curve of the optical transient associated with the GRB \cite{blo99b, gal00} and 
studying the nature of the host galaxies. Of these,
the latter is less direct but relatively simple and robust. 

Associating GRBs with stellar phenomenon necessitates that the host galaxies
are either undergoing active star-formation or have been through epochs 
of substantial star-formation in the past. Since much of the high-redshift star-formation
is obscured by dust, accurately measuring the SFR requires tracing the amount of internal extinction in
the host galaxies. 

Previous star-formation episodes can be constrained either by measuring the metallicity
of the host galaxies or by determining the stellar mass. 
Estimates of galaxy masses derived by fitting template SEDs to the photometry between
UV and visible wavelengths only provide lower limits since they
trace the young stellar component which has a low mass to light ratio.
Accurate masses can only be derived by including
data at rest-frame $\lambda\sim2~\mu$m which traces
the old/cool stellar component. 
At high redshift, IRAC/SIRTF observations will be able to constrain the contribution
from this component.
Thus, by fitting population synthesis models to the rest-frame UV to near-infrared
light one can estimate the mass fraction of gas that has gone through the most recent
starburst, the internal extinction, the age of the starburst and 
fraction of galaxy mass incorporated in an
old stellar population. 

Determining the distribution of stellar ages
in a large number of host galaxies will provide one of the more reliable
ways to distinguish between models where massive stars evolve to become
GRBs. In particular models which involve the production of GRBs 
through the core-collapse of isolated massive stars into a black hole/accretion disk
system would take place on timescales
comparable to the epoch of star-formation i.e.$\sim$10~Myr. On the other 
hand merging double degenerate 
models suffer an evolutionary delay from the epoch of star-formation which is 
induced by the timescale
for orbital decay of the binary system $>$100~Myr.

Thus, if a majority of the GRB hosts
are found to have high specific SFRs (SFR/stellar mass)
and young (age$<$10~Myr) stellar populations,
this would be convincing evidence for `collapsars' to be the progenitors of GRBs.
However, most of the high redshift infrared luminous galaxies
show the presence of a significant older stellar population that is dominating the rest-frame
near-infrared light of the galaxy.
This indicates that infrared luminous galaxies
have undergone previous epochs of star-formation
which could potentially provide an origin for double degenerate systems (Figure 3).
As a result, if the majority of the hosts have a large old stellar population
in addition to the young component, this would make it impossible to discriminate
between the two models using this technique.
A positional analysis of the burst locations on the host 
would then have to used as the discriminator due to the 
large kick velocities imparted to the degenerate objects 
from their supernovae. 
\end{section}


\begin{thebibliography}{10}

\bibitem{berg02}
Berger, E., 2002, these proceedings

\bibitem[Bloom et al. (1999)]{blo99b} Bloom, J. S., et al., 1999, Nature, 401, 453

\bibitem[Chary \& Elbaz (2001)]{cha01} Chary, R., \& Elbaz, D., 2001, ApJ, 556, 
562

\bibitem[Chary, Becklin \& Armus (2002)]{cha02} Chary, R., Becklin, E. E., \& Armus, L., 2002,
ApJ, 566, 1

\bibitem[Elbaz et al. (2002)]{elb02} Elbaz, D., et al., 2002, A\&A, in press (astro-ph/0201328)

\bibitem[Franceschini et al. (2001)]{fra01} Franceschini, A., et al., 2001, A\&A, 378, 1

\bibitem[Galama et al. (2000)]{gal00} Galama, T. J., et al., 2000, ApJ, 536, 185

\bibitem[Galama \& Wijers (2001)]{gal01}
Galama, T. J., \& Wijers, R. A., 2001, ApJ, 549, L209

\bibitem[Meurer et al. (1999)]{meu99} 
Meurer, G. R., Heckman, T. M., Calzetti, D., 1999, ApJ, 521, 64

\bibitem[Ramirez-Ruiz et al. (2002)]{ram01}
Ramirez-Ruiz, E., Trentham, N., \& Blain, A. W., 2002, MNRAS, 329, 465

\end{thebibliography}
\end{document}